\documentstyle[12pt,epsfig]{article}
\newcommand{\la}{$\Lambda$ }
\newcommand{\lab}{$\bar{\Lambda}$ }
\begin{document}
\thispagestyle{empty}
\setlength{\baselineskip}{2.5ex}
\newcommand{\tphi}{\tilde{\phi}}
\newcommand{\lton}{\stackrel{\large <}{\sim}}
\newcommand{\gton}{\stackrel{\large >}{\sim}}
\newcommand{\beq}{\begin{equation}}
\newcommand{\eeq}[1]{\label{#1} \end{equation}}
\newcommand{\beqar}{\begin{eqnarray}}
\newcommand{\eeqar}[1]{\label{#1} \end{eqnarray}}
\newcommand{\gfm}{{\rm GeV/Fm}^3}
\newcommand{\half}{{\textstyle \frac{1}{2}}}
\newcommand{\vx}{{\bf x}}
\newcommand{\vq}{{\bf q}}
\newcommand{\vp}{{\bf p}}
\newcommand{\vk}{{\bf k}}
\newcommand{\vK}{{\bf K}}
\newcommand{\vv}{{\bf v}}
\newcommand{\vb}{{\bf b}}
\newcommand{\vs}{{\bf s}}
\newcommand{\kn}{ $K_0$ }
\newcommand{\knb}{ $\overline{K}_0$ }
\newcommand{\ks}{ $K_s$ }
\begin{center}
{\Large \bf \la and \lab polarization from deep inelastic muon scattering }\\
\end{center}
%
%%%%%%%%%%%%%%%%%%%%%%%%%%%%%%%%%%%%
%%%   E665 Collaboration   %%%%%%%%%
%%%%%%%%%%%%%%%%%%%%%%%%%%%%%%%%%%%%

\vglue0.5cm

%\documentclass[12pt]{article}
% this is based on the rho papre
% still need to add Ashery, Topor Pop
% remove Zeller for Lambda paper
% remove present addresses
%\begin{document}
\def\and{,}
\def\inst#1{$^{#1}$}
\def\author#1{\vskip .5 in #1}
\def\institute#1{\vskip .5 in #1 }
%%%%%%%%%%%%%%%%%%%%%%%%%%%
\noindent
{E665 Collaboration}
\author{
\noindent
M.R. Adams\inst{6} 
\and M. Aderholz\inst{12}
\and S. A\"{\i}d\inst{10}
\and P.L. Anthony\inst{9}
\and D. Ashery\inst{2,18}
\and D.A. Averill\inst{6}
\and M.D. Baker\inst{11}\and \\
B.R. Baller\inst{4}
\and A. Banerjee\inst{15} 
\and A.A. Bhatti\inst{16}
\and U. Bratzler\inst{16} 
\and H.M. Braun\inst{17}
\and T.J. Carroll\inst{12}
\and H.L. Clark\inst{14}\and \\
J.M. Conrad\inst{5}
\and R. Davisson\inst{16} 
\and I. Derado\inst{12}
\and F.S. Dietrich\inst{9} 
\and W. Dougherty\inst{16} 
\and T. Dreyer\inst{1} 
\and V. Eckardt\inst{12}\and \\
U. Ecker\inst{17}
\and M. Erdmann\inst{1}
\and G.Y. Fang\inst{5}
\and J. Figiel\inst{8} 
\and R.W. Finlay\inst{14} 
\and H.J. Gebauer\inst{12} 
\and D.F. Geesaman\inst{2}\and\\
K.A. Griffioen\inst{15}
\and R.S. Guo\inst{6}
\and J. Haas\inst{1} 
\and C. Halliwell\inst{6} 
\and D. Hantke\inst{12}
\and K.H. Hicks\inst{14} 
\and H.E. Jackson\inst{2}\and \\
%D.E. Jaffe\inst{6}
G. Jancso\inst{7}\and 
D.M. Jansen\inst{16}
\and Z. Jin\inst{16}
\and K. Kadija\inst{12}
\and S. Kaufmann\inst{2} 
\and R.D. Kennedy\inst{3}\and 
E.R. Kinney\inst{2}\and \\
H.G.E. Kobrak\inst{3} 
\and A.V. Kotwal\inst{5}
\and S. Kunori\inst{10} 
\and M. Lenski\inst{1}
\and J.J. Lord\inst{16}\and 
H.J. Lubatti\inst{16} 
\and D. McLeod\inst{6} \and\\
A. Manz\inst{12} 
\and H. Melanson\inst{4} 
\and D.G. Michael\inst{5} 
\and H.E. Montgomery\inst{4} 
\and J.G. Morfin\inst{4}\and 
R.B. Nickerson\inst{5}\and \\
 K. Olkiewicz\inst{8} 
\and L. Osborne\inst{11}
\and R. Otten\inst{17} 
\and V. Papavassiliou\inst{2}\and 
B. Pawlik\inst{8}\and 
F.M. Pipkin\inst{\dag 5}\and\\
D.H. Potterveld\inst{2} 
\and A. R\"oser\inst{17} 
\and J.J. Ryan\inst{11} 
\and C.W. Salgado\inst{4} 
\and H. Schellman\inst{13} 
\and M. Schmitt\inst{5} 
\and N. Schmitz\inst{12}\and \\
G. Siegert\inst{1}
\and A. Skuja\inst{10} 
\and G.A. Snow\inst{10} 
\and S. S\"oldner-Rembold\inst{12}
\and P. Spentzouris\inst{4}
\and P. Stopa\inst{8}\and\\
 R.A. Swanson\inst{3}
\and  V. Topor Pop\inst{18}
\and H. Venkataramania\inst{13} 
\and M. Wilhelm\inst{1}
\and Richard Wilson\inst{5} 
\and W. Wittek\inst{12}\and\\
S.A. Wolbers\inst{4}\and  %\and G.P Zeller\inst{13}
A. Zghiche\inst{2} 
\and T. Zhao\inst{16}
}
\def\and{\\}
\institute{
\noindent
{\it{$^{1}$Albert-Ludwigs-Universit\"at Freiburg i.~Br., Germany }} \and
{\it{$^{2}$Argonne National Laboratory, Argonne, Illinois 60439}} \and
{\it{$^{3}$University of California, San Diego, California 92093}} \and
{\it{$^{4}$Fermi National Accelerator Laboratory, Batavia, Illinois 60510}} \and
{\it{$^{5}$Harvard University, Cambridge, Massachusetts 02138}} \and
{\it{$^{6}$University of Illinois, Chicago, Illinois 60680}} \and
{\it{$^{7}$KFKI Research Institute for Particle and Nuclear Physics,
H-1525 Budapest, Hungary }}\and
{\it{$^{8}$Institute for Nuclear Physics, Krakow, Poland}} \and
{\it{$^{9}$Lawrence Livermore National Laboratory, Livermore, California}} 94551\and
{\it{$^{10}$University of Maryland, College Park, Maryland 20742}} \and
{\it{$^{11}$Massachusetts Institute of Technology, Cambridge, Massachusetts 02139}} \and
{\it{$^{12}$Max-Planck-Institut f\"{u}r Physik, Munich, Germany}} \and
{\it{$^{13}$Northwestern University, Evanston, Illinois 60208}} \and
{\it{$^{14}$Ohio University, Athens, Ohio 45701}} \and
{\it{$^{15}$University of Pennsylvania, Philadelphia, Pennsylvania 19104 }}\and
{\it{$^{16}$University of Washington, Seattle, Washington 98195}} \and
{\it{$^{17}$University of Wuppertal, Wuppertal, Germany}} \and
{\it{$^{18}$Tel Aviv University, Tel Aviv, Israel}}}\\

{\small
\noindent
$^{\dag}$ deceased \\
}
%%%%%%%%%%%%%%

%\newpage

\begin{abstract}
We report results of the first measurements of \la and \lab polarization
produced in deep inelastic polarized muon scattering on the nucleon. The 
results are consistent with an expected trend towards positive polarization with increasing 
%of increasing positive polarization for large 
$x_{\rm F}$. The polarizations of \la and \lab appear to have opposite
signs. A large negative polarization for \la at low positive $x_{\rm F}$ is 
observed and is not explained by existing models. A possible interpretation is
presented.
\end{abstract}\
%\newpage

The spin structure function of the nucleon has been studied extensively 
during the past several years. Very different experimental systems utilizing
polarized muon \cite{smc}, electron \cite{slac} and positron \cite{herm}
beams and covering different kinematic regions obtained results which are
consistent with each other. This effort led to precise measurements of the 
nucleon spin structure functions $g_1^N(x)$ (N = n,p) and their integrals
$\Gamma_1^N = \int{g_1^N(x)dx}$. 
Here $x \equiv  x_{\rm Bj} = Q^2/2M\nu$ is the Bjorken scaling variable, $-Q^2$
is the four-momentum transfer to the target nucleon squared, $\nu$ is 
the energy loss of the lepton in the laboratory frame, and $M$ is the 
nucleon mass.
The interpretation of these results is 
that quarks in the 
nucleon carry only $\sim$30\% of the nucleon spin and that the strange
(and non-strange) sea is polarized opposite to the polarization of the valence
quarks. This interpretation leads to the question of where the rest of
the nucleon spin is coming from which will not be addressed here.
Other open questions are: 1) how sensitive
are the conclusions to the SU(3) flavor symmetry assumed in the 
interpretation of the experimental data? 2) what is the mechanism that 
polarizes the strange sea? 3) how well do we understand the spin structure
of other hadrons? 4) are the quarks and antiquarks in the sea equally
polarized? \\

%These questions can be addressed through 
We have addressed these questions with a measurement of the polarization
of \la and \lab hyperons produced in polarized muon Deep Inelastic 
Scattering (DIS) on unpolarized targets. 
Having the strange quark as valence rather than sea quark, the
polarization of \la hyperons is sensitive to its contribution to the
\la spin. By measuring both \la and \lab polarization the roles of both
$s$ and $\bar{s}$ can be studied. Models exist that describe
mechanisms of the strange sea polarization \cite{ekks,elliskk} and 
predict the \la polarization in the target fragmentation region. 
Other models are based on the na\"{\i}ve quark model or invoke SU(3) 
symmetry to calculate the spin structure function of the \la from that of 
the nucleon \cite{koz,burjaf,jaf,def,thom,dazl}. These models predict the
\la 
and \lab polarization in the current fragmentation region.
These predictions can be tested by measuring the polarization
of the \la and \lab.\\

As an illustration, a simple mechanism that can produce \la polarization in 
polarized lepton DIS is the following: the polarized virtual 
photon is absorbed by a strange quark in the target 
nucleon sea. The struck quark then emerges with spin aligned in
the direction of that of the photon. When it hadronizes into a \la (in the
current fragmentation region), it is likely to become a valence quark
in the \la and the
na\"{\i}ve quark model predicts that the polarization of the \la will be
the same as that of the strange quark. 
An experiment that can probe the low $x_{\rm Bj}$ region will have a significant
fraction of the \la and \lab originating from this mechanism and will 
therefore have sensitivity to the strange sea in the target nucleon. 
When the DIS kinematics selects the sea quarks (low $x_{\rm Bj}$) the 
probability that the struck quark will be a strange quark is about an
order of magnitude smaller than for it to be a $u$ or $d$ quark. This is
because the scattering probability is proportional to the electric charge 
squared and there are half as many strange quarks in the sea as there are 
$u$ and $d$ quarks. However, if $u$ or $d$ quarks are struck out their
probability
to hadronize into a strange baryon is small \cite{det}. On the other hand,
if a strange quark is struck out it will {\em always} hadronize into a
strange particle including baryons. It can therefore be expected
that a significant fraction of leading strange baryons will carry the
strange struck quark.\\

The transfer of polarization from the beam to the produced \la and \lab
is controlled by their helicity difference fragmentation functions
$\Delta \hat{q}_{\Lambda}$. 'Complexities' in the analysis of fragment
polarization, as defined in reference \cite{jaf}, are proportional to sin$^2\theta_{\mu}$
where $\theta_{\mu}$ is the laboratory scattering angle of the muon.
%The
%distribution of this angle for E665 requiring detection of a \la is hown
%in fig. \ref{fig:tetmu}. It can be seen that the corrections are of
%the order of $< ~10^{-4}$. Under these conditions
If these are small, as they are in this experiment, the measured polarization of a \la produced in DIS
of muons having polarization $P_{\mu}$ is given by:

\begin{equation}
\vec{P}^{meas}_{\Lambda} = \vec{P}_{\mu}D(y)
\frac{\sum_q e_q^2 q_N(x,Q^2)\Delta \hat{q}_{\Lambda}(z,Q^2)}
{\sum_q e_q^2 q_N(x,Q^2)\hat{q}_{\Lambda}(z,Q^2)}.
\label{eq1}
\end{equation}

Here $D(y) = y(2 - y)/[1 + (1 - y)^2]$, $y = \nu/E_{\mu}$, $e_q$ and
$q_N$ are the charge and $x$-distribution of the quark with flavor $q$, 
respectively, and $z$ is the energy of the produced hadron divided by $\nu$.
In recent measurements at LEP $\Delta \hat{q}$ was measured from \la 
polarization near the Z pole and was found to be consistent with the
na\"{\i}ve quark model expectations \cite{aleph,opal}. \\
%Other mechanisms exist 
%and are discussed in details in various theoretical papers 
%\cite{ekks,elliskk,burjaf,jaf,def,thom,dazl}\\

In this paper we present the first results of \la and \lab polarization
measured in DIS with polarized muon beams. These results were obtained from data 
taken by the Fermilab E665 collaboration. In the experiment
a 470 GeV/c positive muon beam was scattered off hydrogen, deuterium and
several nuclear targets. The scattered muons and the final state hadrons were
reconstructed in an open-geometry double-dipole spectrometer which had 
acceptance for forward produced charged hadrons. 
A more complete description of the experimental system can be found
elsewhere \cite{mc3}. 
The muon beam, which comes from $\pi$ and $K$ decays is naturally polarized at
a level determined by the beam optics.
The polarization of the muon beam 
was calculated to be P$_{\mu} = -0.7 ~\pm$ 0.1. The analysis presented here covers
$10^{-4} < x_{\rm Bj} < 10^{-1}$ with $\langle x_{\rm Bj}\rangle = 5\cdot 10^{-3}$,
$0.25 < Q^2 < 2.5 ~\rm {GeV}^2/c^2$ with $\langle Q^2 \rangle = 1.3  
~\rm {GeV}^2/c^2$ and $\langle \nu \rangle = 150$ GeV.  The values of
sin$^2\theta_{\mu}$ are less than $10^{-4}$ which satisfies the above mentioned
conditions. The data was taken with unpolarized hydrogen 
and deuterium targets.\\

The \la and \lab were identified by applying a kinematic fit 
to pairs of a positive and a negative track and reconstructing their
invariant mass. The fit exploited the different kinematics that
characterize the $\Lambda \rightarrow \pi p$ and 
$K^0 \rightarrow \pi^+ \pi^-$ decays and selected events likely to be the
former.  This and the requirement that the decay vertex is downstream, 
well separated from the production vertex and pointing to it 
led to the selection of a very clean signal. 
Background effects were studied by varying the \la and \lab selection
criteria, in particular, by replacing the kinematic fit by other requirements
such as inconsistency with the $K^0$ mass if both tracks are assigned the
pion mass. These criteria selected signals which left some background 
below and above the \la mass peak. Using these two signals, and
after subtracting the distributions of the background, 
we produced two cos$\theta_{p,\gamma^*}$ distributions as used 
to deduce the polarization (see below).
The two distributions were found to be consistent with 
each other. Contributions of the background to the systematic uncertainties 
in the polarization were taken from comparison of these distributions.
Secondary interactions were found to account for less than 3\% of the
events.\\

For the final sample we also 
required: $m_{e^+e^-} > $ 0.05 GeV, $\nu > $ 50 GeV, $ 0.1 < y < 0.8$ and
$p_{\pi} > 4$ GeV/c. Here  $m_{e^+e^-}$ is the invariant mass of the track pair
assuming that both tracks are electrons and $p_{\pi}$ is the pion momentum.
The size of the sample with the final cuts was about
750 \la and 650 \lab hyperons. The similarity  of the \la and \lab yields 
is an important feature of this experiment and results from the low $x_{\rm Bj}$
sensitivity probing mainly the sea quarks of the target nucleon. The
small difference may be caused by \la hyperons produced by 
contribution from valence quarks in this $x_{\rm Bj}$ region.
The size of the sample does not allow a detailed study of the $x_{\rm F}$ 
dependence of the polarization ($x_{\rm F} = p_l/p_{l ~max}$ is the 
Feynman-x variable, $p_l$ is the longitudinal
momentum relative to the virtual-photon direction and $p_{l ~max}$ is
the maximal value it can have, both in the hadronic c.m. system).
We divided the sample into two $x_{\rm F}$ bins: 
$0 < x_{\rm F} < 0.3$ with $\langle x_{\rm F} \rangle$ = 0.15 and
$0.3 < x_{\rm F} < 1.0$ with $\langle x_{\rm F} \rangle$ = 0.44. We also present
results for $x_{\rm F} > 0.1$. \\

\begin{figure}[p]
%\centerline{\epsfbox{draft_fig_1_2.ps}}
\epsfig{figure=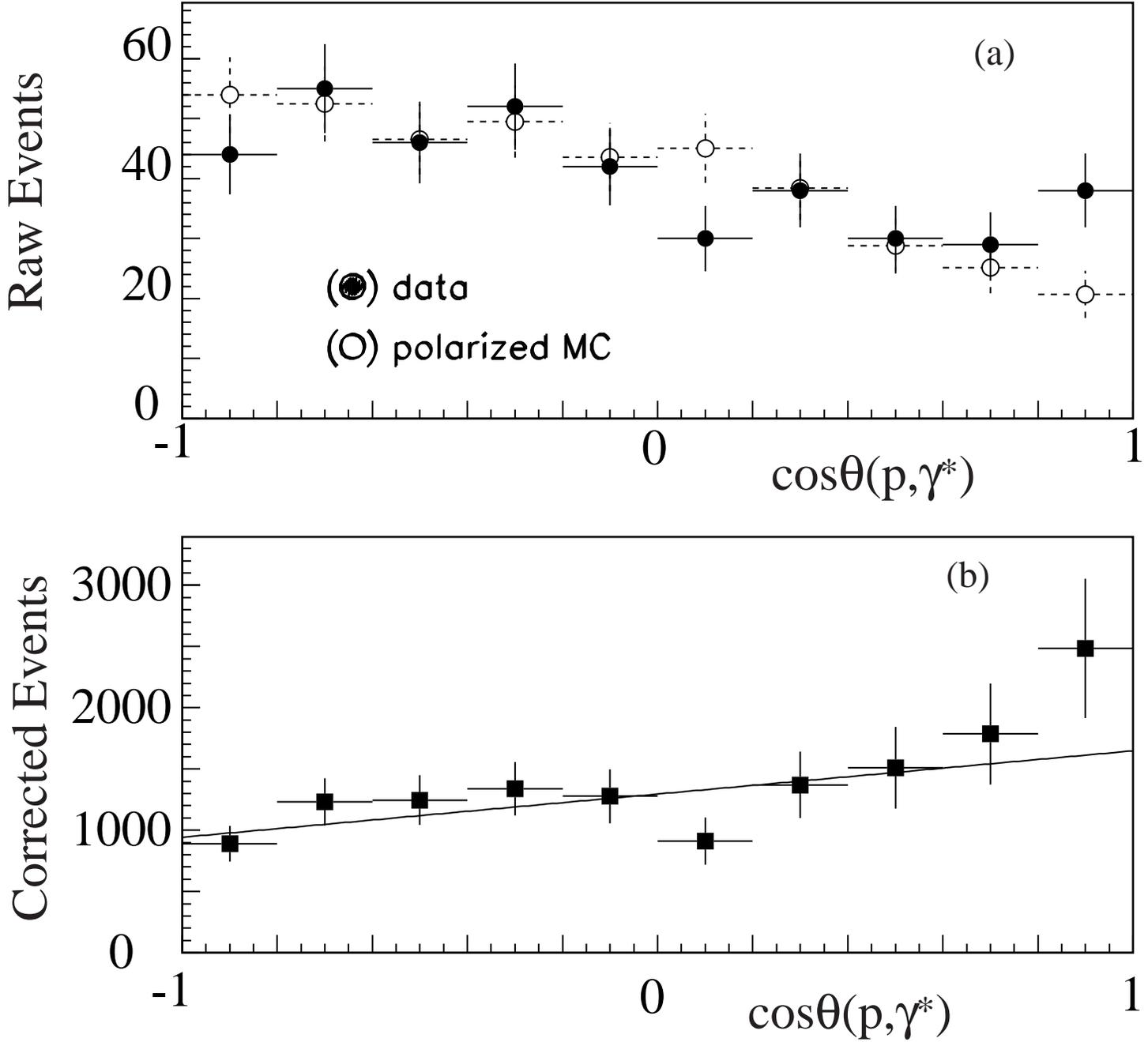}
\caption{cos$\theta_{p,\gamma^*}$ distributions for 
\la in the $0 < x_{\rm F} < 0.3$ region. (a) 
Raw data distribution and distribution of the polarized Monte Carlo 
simulation (see text).
(b) Data corrected for experimental acceptance and the linear fit 
($\chi^2/dof$ = 1.2 ) used to deduce the polarization. Shown are statistical
errors only.}
\label{fig1}
\end{figure}

The angular distribution $\theta_{p,\gamma^*}$ 
of the proton in the $\Lambda \rightarrow \pi p$
decay is computed in the \la rest frame with respect to the direction
of the absorbed virtual photon. The measured angular distribution is shown
in figure \ref{fig1}(a). This decay angular distribution is affected 
by detector acceptance. In order to correct for that we carried out
extensive 
Monte Carlo simulations of the decay distributions applying the same analysis
cuts as for the data. 
        The primary $\mu p$ interaction is simulated 
        using the standard LUND event generator (LEPTO 5.2,
        JETSET 6.3 \cite{mc1}). Initial and final state radiative
        events are generated with the GAMRAD \cite{mc2} Monte Carlo. 
        Particles produced at the interaction vertex 
        are tracked through the E665 spectometer \cite{mc3} using 
        the GEANT \cite{mc4} program.  A detailed description of 
        the Monte Carlo simulation has been documented
        elsewhere \cite{mc5}.
% only those aspects which are relevant to this analysis will be discussed.
These simulations were found to reproduce well 
the distributions of many observables in this
experiment and in the present analysis. In particular, good reproduction of
the pion momentum distribution, a crucial parameter in this analysis, was
observed for $p_{\pi} > 3$ GeV/c. \\

The acceptance corrections were taken as the ratio of the number of 
generated events  and the number of
reconstructed events accepted by the selection criteria as applied to the
data. In order to test possible background
effects on these corrections we used two separate Monte Carlo simulations,
one in which the ratio of generated $K^0/\Lambda$ was about 12 and another
in which it was about 0.6. In spite of the factor 20 difference in the ratio 
the two simulations yielded the same correction factors. The comparison
was used to estimate the contribution from background in the simulations
to the systematic uncertainties in the acceptance corrections.
The acceptance-corrected angular distributions are expected to have the form:

\begin{equation}
\label{eq:lpol}
I(\theta) = 1 + \alpha P^{meas}_{\Lambda} \cos \theta_{p,\gamma^*}
\end{equation}
where $P^{meas}_{\Lambda}$ is the measured \la polarization and the value of the 
asymmetry parameter $\alpha$ has been
determined experimentally \cite{pdg} to be $0.642 ~\pm$ 0.013 ($-0.642$ for
\lab). The experimental resolution in
cos$\theta_{p,\gamma^*}$ was found to be about 0.02, small compared with
the bin size. The acceptance-corrected angular distributions were fitted to 
straight lines. Such a distribution is shown in figure \ref{fig1}(b). 
We deduce the polarization of the \la and \lab from the
slopes of the fits. 
After obtaining the polarizations we repeated the Monte Carlo simulations,
this time introducing the measured polarization in the Monte Carlo
generated events. After analysing the reconstructed events from the 
polarized Monte Carlo in the same way as the data the resulting 
distributions agreed well with those of the data (fig. \ref{fig1}(a)).\\

The systematic uncertainties in the results are due to uncertainties in
the {\em slopes} of the aceptance corrections (not the absolute corrections).
These were estimated from comparison of the slopes using the different
selection criteria and using the various simulations: with different
$K^0/\Lambda$ ratios and with polarized and unpolarized generated 
distributions. The
contributions from these sources were added in quadrature and resulted in
a systematic uncertainty in the polarization of $\pm$0.1, for both 
$x_{\rm F}$ bins.\\

In this analysis cos$\theta_{p,\gamma^*}$ is measured with respect to the 
virtual photon momentum. However, the interesting reference direction 
is that of the virtual photon spin. 
Since in this experiment the polarization of the muon beam and hence that
of the virtual photon are negative we apply a minus sign to the results of the
preceeding analysis.
The resulting measured polarizations are summarized in table \ref{ta:fin}. The
total error $\Delta P^{meas}_{\Lambda}$ is obtained by adding the systematic and statistical
errors in quadrature. Finally, by dividing-out the diluting effects of the beam
polarization $| P_{\mu} |$ = 0.7 $\pm$ 0.1 and the photon 
depolarization factor $D(y)$ which lies in the range 0.4 to 0.5, we
obtain the undiluted polarizations P$_{\Lambda}$ which are listed in table \ref{ta:fin}.

\newpage

\begin{table}[H]
\caption{Longitudinal polarizations $\bf{P^{meas}_{\Lambda}}$ obtained from straight
line fits to cos$\theta_{p,\gamma^*}$ distributions of data
corrected for detection acceptance and the undiluted polarizations $\bf{P_{\Lambda}}$.
 The signs are with respect to the 
direction of the spin of the virtual photon.
The results are shown for \la and \lab in various $x_{\rm F}$ ranges.}
\vskip 1.0cm
\centering
 \begin{tabular}{|c|c|c|c|c|c|c|c|} \hline\hline
&$x_{\rm F}$ range& $\langle x_{\rm F} \rangle$  & $P^{meas}_{\Lambda}$ &
 $\Delta P_{stat}$  & $\Delta P^{meas}_{\Lambda}$ & $P_{\Lambda}$ & $\Delta P_{\Lambda}$  \\
\hline\hline\hline\hline
  &  &  &  &  &  &  &  \\
\la & 0. - 0.3 & 0.15 & $-$0.42  & 0.14 & 0.17 & $-$1.2 & 0.5 \\
  &  &  &  &  &  &  &\\
\la & 0.3 - 1.0 & 0.44 & $-$0.09  & 0.16 & 0.19 & $-$0.32 & 0.7\\
  &  &  &  &  &   &  &\\
\la & 0.1 - 1.0 & 0.31 & $-$0.23  & 0.07 & 0.12 &  $-$0.74 & 0.4 \\
  &  &  &  &  &  &  & \\
\hline\hline\hline\hline
  &  &  &  &  &  &  & \\
\lab & 0. - 0.3 & 0.15 & 0.09  & 0.17 & 0.20 & 0.26 & 0.6\\
  &  &  &  &  &   &  &\\
\lab & 0.3 - 1.0 & 0.44 & 0.31  & 0.20 & 0.22 & 1.1 & 0.8 \\
   &  &  &  &  &   &  & \\
\lab & 0.1 - 1.0 & 0.31 & 0.02  & 0.13 & 0.16 & 0.06 & 0.5 \\
  &  &  &  &  &  &  & \\
\hline\hline\hline\hline
\end{tabular}
\label{ta:fin}
\end{table}

\begin{figure}[t]
%\centerline{\epsfbox{/silly2/ashery/pol/mc_study/pol_expect_ex_1.ps}}
\epsfig{figure=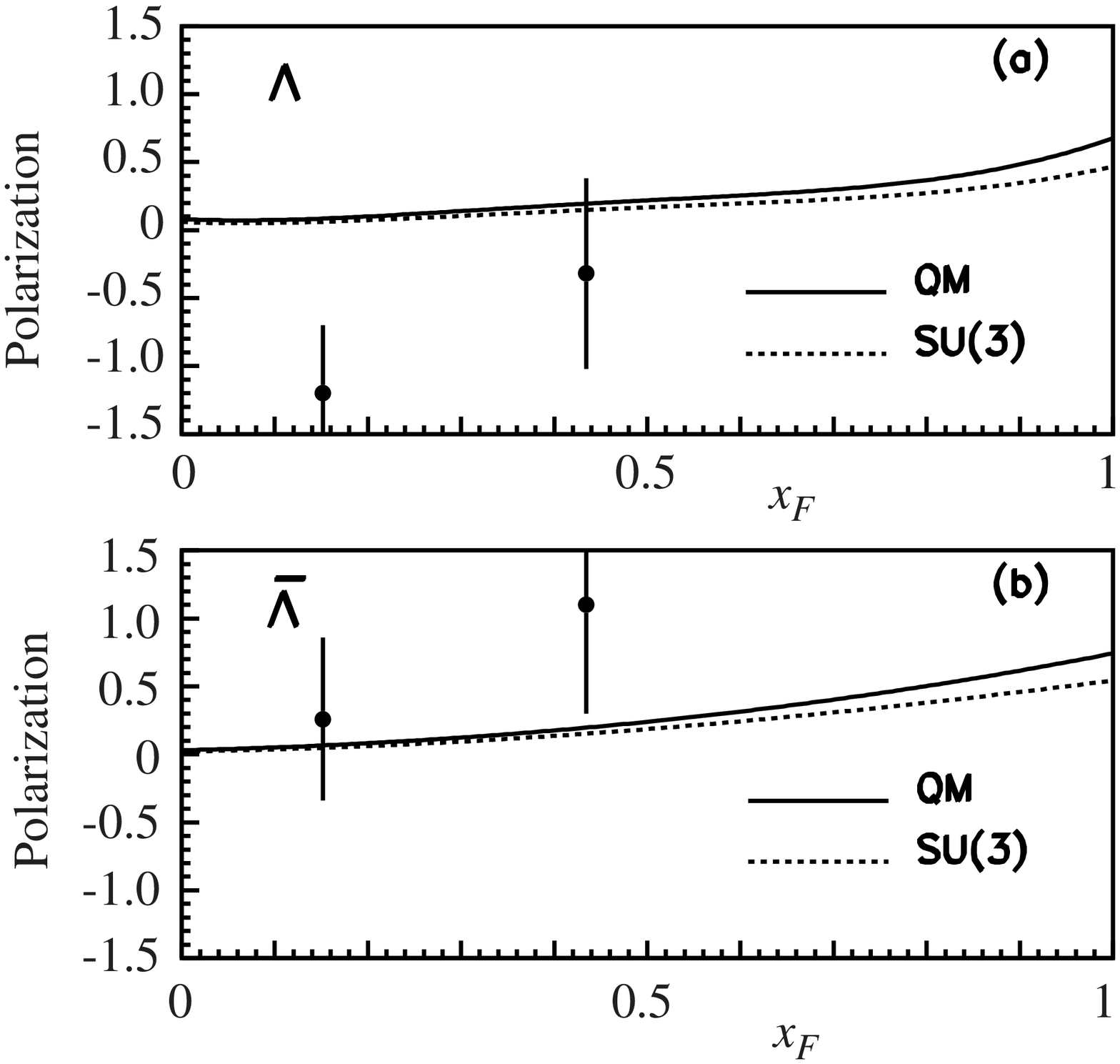}
\caption{The polarization $P_{\Lambda}$ of \la and \lab hyperons. The full circles 
are for data in the low and high $x_{\rm F}$ bins corrected for beam polarization and
photon depolarization. The curves are from \protect\cite{dazl} and show
the
predictions of the na\"{\i}ve quark model and by assuming SU(3) symmetry. }
\label{comp}
\end{figure}

In the present analysis we cannot distinguish between \la and \lab
produced directly from hadronization of a struck quark 
and \la and \lab that are decay products. The main contributing decays are 
$\Sigma^0 \rightarrow \Lambda \gamma$ and $\Sigma^* \rightarrow
\Lambda\pi$. Following the procedure adopted by \cite{aleph,opal} we use
the Monte Carlo simulations discussed above to estimate the contributions
from all sources.
In figure \ref{comp} we compare the experimental results with calculations
based on two models \cite{dazl} pertaining to \la and \lab detected in the 
current fragmentation region. One is the na\"{\i}ve quark model where all 
baryons are three-quark states with wave functions having zero orbital angular 
momentum and all the spin of the baryon comes from quark spins. The second is
the SU(3)$_{\rm F}$ symmetry model where the spin structure of SU(3) octet hyperons
can be deduced from that of the proton \cite{jaf}. 
%The third model, in which SU(3)$_{\rm F}$ symmetry breaking is introduced 
%\cite{zvi,jezvi}, yields results indistiguishable from those of the second. 
The calculations shown by the curves in figure \ref{comp} were done 
according to the procedures
derived in \cite{dazl} and adapted to the experimental conditions of the
present data. The E665 Monte Carlo is used to simulate the contributions
from struck quarks and from $\Sigma^0$ and $\Sigma^*$ decays. 
%as well as all other experimental effects. 
The polarizations 
%include the absolute value of the muon beam polarization (0.7 $\pm$ 0.1)
%and the photon depolarization factor $D(y)$ (0.4 - 0.5). They 
were computed with respect to the direction of the virtual-photon spin.\\

Measurements of \la polarization at the Z pole \cite{aleph,opal} show good agreement
with calculations in which the spin structure of the \la was taken from the 
na\"{\i}ve quark model
and the fragmentation calculated using JETSET-based Monte Carlo
simulations. This
agreement motivated the calculations shown in figure 2 as discussed above
\cite{dazl}. These calculations show a trend of increasing
positive polarization with increasing $x_{\rm F}$ that seems to be suggested
by the data. The statistics and $x_{\rm F}$
range of the present data do not allow to confirm this trend more precisely.
The large negative value of the measured \la polarization 
at $\langle x_{\rm F} \rangle $ = 0.15
is intriguing. For such low $x_{\rm F}$ values the models shown in figure 2
are not very reliable as these models are expected to work better for 
relatively large $x_{\rm F}$. This result suggests a significant contribution from target 
fragmentation effects which appear at low and negative $x_{\rm F}$ values and favor
the \la over the \lab \cite{emc}. In anti-neutrino data a large
negative \la polarization has also been observed \cite{nu}. Ellis {\em et al.}
\cite{ekks,elliskk} have been able to explain this result by assuming
that all the sea quark pairs are coupled to a triplet-p state ($^3P_0$)  and are 
polarized oppositely to the valence quark. The authors apply their model to Muon 
DIS only for the case where the virtual photon is absorbed by a valence
quark. In this case, which is not appropriate for the kinematics of the present work,
they predict a positive \la polarization. However, their model can be extended to the
case where the virtual photon is absorbed by a sea quark as the kinematics of the 
present work imply.
If we assume that all $s\bar{s}$ pairs in the sea are coupled to $^3P_0$,
when a positively polarized photon
is absorbed by the negatively polarized member of the pair, the remnant
strange quark is likely to be negatively polarized. If the spin of the 
remnant \la is determined by that of the remnant strange quark, 
it will also have a negative polarization.
While the actual process may be more complex and these arguments cannot
describe it quantitatively, they can indicate a possible source for negative 
polarization as is actually observed.\\

In conclusion, we have presented the first results of \la and
\lab polarization measured in polarized deep inelastic muon
scattering on the nucleon. The results are consistent with the
expected trend towards positive polarization with increasing $x_{\rm F}$.
A large negative polarization of the \la at low $x_{\rm F}$ is observed
suggesting that target fragmentation remains important in this region.

\vspace{1cm}
This work was performed at the Fermi National Accelerator Laboratory, which
is operated by Universities Research Association, Inc., under contract
with the U.S. Dept. of Energy.  The work was supported
by the U.S Department of Energy, the  National Science Foundation (USA), the
Bundesministerium f\"ur Forschung und Technologie (Germany), the Polish
Committee for Scientific Research, the Hungarian Science Foundation and
the Israel Science Foundation.

%\newpage
%\thispagestyle{empty}

\bibliographystyle{unsrt}

\end{document}